# Optical Orientation in Ferromagnet/Semiconductor Hybrids

**V.L. Korenev**

A.F. Ioffe Physical Technical Institute, Russian Academy of Sciences, 194021, Saint-Petersburg, Russia

**Abstract.** The physics of optical pumping of semiconductor electrons in the ferromagnet/semiconductor hybrids is discussed. Optically oriented semiconductor electrons detect the magnetic state of the ferromagnetic film. In turn, the ferromagnetism of the hybrid can be controlled optically with the help of the semiconductor. Spin-spin interactions near the interface ferromagnet/semiconductor play crucial role in the optical readout and the manipulation of ferromagnetism.

**1. Introduction**

Nowadays large efforts are directed to the integration of magnetism into semiconductor architecture of modern computers. The first approach is to construct a universal object combining ferromagnetic (FM) and semiconducting (SC) properties. Following pioneering research of the 60-s of the last century on Eu-based chalcogenides and Cd-Cr spinels, Ohno (1998) and Dietl *et al* (2000) initiated the recent active studies of new III-V-based ferromagnetic semiconductors. The second approach – the so-called "semiconductor spintronics" – is based on generation and manipulation of spin and spin currents entirely within non-magnetic semiconductors (Zutic *et al* 2004). The third approach (Prinz 1990) exploits ferromagnet/semiconductor (FM/SC) hybrid systems. One of the advantages is an additional degree of freedom consisting in the choice of desirable pair FM/SC among paramagnet semiconductors and a large number of ferromagnetic materials. As demonstrated by Zakharchenya and Korenev (2005), another advantage is the possibility both to read and to control the magnetization of FM with the help of semiconductor. Here I focus on the physics of spin interactions in FM/SC hybrids and their effect on the optical orientation of spins in semiconductor.

Electron spin polarization is sensitive monitor of various spin interactions. For example, a ferromagnetic film creates a static magnetic field. Although relatively weak, this magnetic field penetrates deep into semiconductor. The electron spin interacts with the stray magnetic field by means of Zeeman interaction. Randomly distributed stray fields bring about the Larmor precession of semiconductor electron spins, thus leading to their spin relaxation. This effect (section 2) can be used to distinguish between magnetized and demagnetized states of the ferromagnet.

The exchange interaction (Coulomb interaction restricted by the Pauli exclusion principle) of charge carriers is the strongest spin interaction in hybrids. It is short-range depending strongly on the wavefunction overlapping. The exchange interaction causes ferromagnetism and splits the energy levels of the ferromagnet itself by a large value (~1 eV). The wavefunction overlap between ferromagnetic atoms and semiconductor charge carriers in a FM/SC hybrid leads in equilibrium to the local spin polarization of semiconductor carriers – the so-called proximity effect. In non-equilibrium conditions the FM-induced spin polarization is transferred into semiconductor over the macroscopic distance (~10 μm) leading to the dynamical proximity effect (section 3).

The FM-SC interfacial exchange coupling forms a unified spin system whose properties differ substantially from those of an isolated FM film. At first glance this sounds strange because the density of charge (and spin) carriers in paramagnetic semiconductors is much less than the density of magnetic atoms in typical ferromagnets. Therefore, "common sense" suggests that a semiconductor is not capable to manipulate the powerful ferromagnetic spin system. However, the contact of FM film and SC leads to the band bending (Schottky barrier) and accumulation of a fair quantity of charge carriers (electrons or holes) near the interface. The strong exchange interaction of SC charge carriers and magnetic atoms of FM leads to a strongly coupled spin system. The uniqueness of the FM/SC system lies in the electrical and optical tunability of electrical and magnetic properties of the paramagnetic SC. This means that the ferromagnetism (for example, magnetic hysteresis loop) of the unified system can be tuned as well. As a result, SC is not only a substrate for a FM to lie upon but participates *actively* in information processing (section 4).

## 2. Zeeman interaction in FM/SC hybrid

An electron with spin $\vec{s}$ and magnetic moment $\vec{\mu} = -\mu_B g \vec{s}$ interacts with an external magnetic field $\vec{H}$. The Hamiltonian of Zeeman interaction

$$\hat{H}_Z = -\vec{\mu}\cdot\vec{H} = \mu_B g \vec{s}\cdot\vec{H},$$

where $\mu_B = \dfrac{|e|\hbar}{2m_e c} \approx 60\,\mu eV/T$ is the Bohr magnetron, and $g$ is the electron's $g$-factor. The interaction leads to spin precession around $\vec{H}$ with a Larmor frequency, $\vec{\Omega} = \mu_B g \vec{H}/\hbar$ according to the equation of motion $\dot{\vec{s}} = [\vec{\Omega}\times\vec{s}]$. The spin precession of particles is also affected by local magnetic fields of other sources, and in general the total field is the vector sum of the fields.

### 2.1. *Magnetostatic stray fields created by the ferromagnetic film in semiconductor*

Local stray magnetic fields generated by ferromagnetic film affect the semiconductor electron spin similarly to the external magnetic field. The distribution of stray fields in space depends strongly on the magnetic state of the FM film. The magnetic field outside the uniformly magnetized laterally infinite film is zero because the magnetic lines of force close at infinity. The film of finite lateral size $L$ creates at point A (Fig.1a) a magnetic field $\vec{H}_A \approx 4\pi\vec{M}\,d/L$ ($d$ is the film thickness, $\vec{M}$ is magnetization). It penetrates deep $\sim L$ into SC. For a Ni film with $M = 510\,Oe$, $d = 10$ nm and $L=0.1$ cm the field $H_A \approx 0.06$ Oe. The stray field near the FM/SC interface is the strongest when the film is demagnetized, i.e. broken into a large number of domains with different orientations of magnetization. In this case both the direction and strength of magnetic field vary in space with characteristic scale determined by the domain size $\ell$ (Fig.1b). For $\ell = 10\,\mu$ the field $H_A \sim 6\,Oe$ is hundred times larger than that of the uniformly magnetized FM. However it decays depthward much faster – over the length $\geq \ell$.

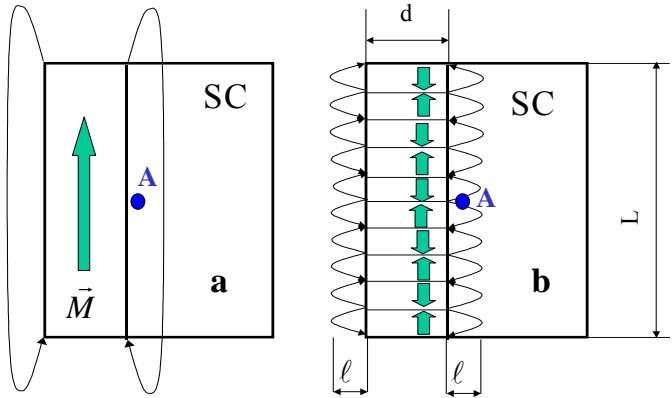

Figure 1. Stray magnetic field created by the FM film

*2.2 Effect of stray fields on the optical orientation of semiconductor electrons.*

Dzhioev et al (1994, 1995) observed the spin precession of SC electrons in static stray fields in a Ni/n-GaAs hybrid by optical orientation method. The essence of optical orientation (Meier and Zakharchenya Eds, 1984) is the generation of spin-polarized conduction band electrons by circularly polarized light transferring the photon angular momentum into the spin system of semiconductor electrons. In its turn, annihilation of the electrons emits circularly polarized light. The degree $\rho$ of the circular polarization of photoluminescence (PL) is equal to the projection of the electron ensemble-averaged spin $\vec{S}_c$ onto registration direction z, usually coinciding with the normal to the structure plane (top of Figure 2). An external magnetic field induces precession of each electron spin about $\vec{H}$ with Larmor frequency $\vec{\Omega}$. In steady state conditions, however, the average spin does not change with time but deflects from the initial direction (z) decreasing in absolute value (Fig.2). This leads to the Hanle effect – depolarization of PL with magnetic field. The halfwidth $H_{1/2}$ of the magnetic depolarization curve is determined by the equality of frequency $\mu_B g H_{1/2}/\hbar$ and reciprocal non-equilibrium spin lifetime $1/T_s$. The longer the spin lifetime $T_s$, the smaller magnetic field necessary to rotate spin and depolarize PL. As a result, we have the Hanle effect-based optical magnetometer whose sensitivity is determined by the $H_{1/2}$ value. As we discussed in subsection 2.1, the amplitude of the stray fields is pretty small (~ 1 G), so that the halfwidth should be narrow enough. Such a narrow halfwidth is realized in n-GaAs samples (see Chapter by Kavokin and Dzhioev). Figure 3a compares the Hanle effects in n-GaAs when the film is previously magnetized (filled circles) and demagnetized (open circles). It is seen that the demagnetization decreases the degree $\rho$ with the maximum difference being achieved in the zero external magnetic field.

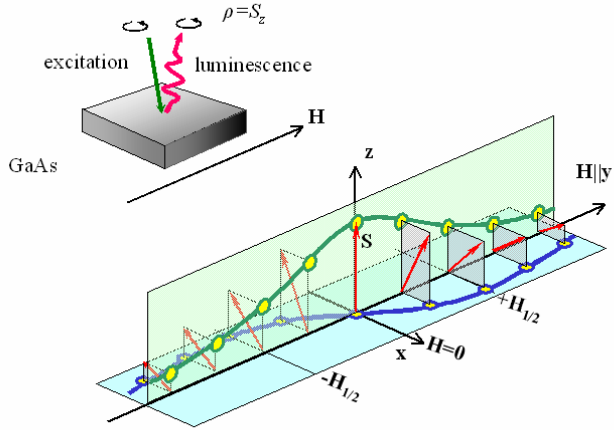

Figure 2. Optical orientation experiment and the Hanle effect of semiconductor electrons. Arrows show the steady state mean electron spin in transversal magnetic field of different values. (Adapted from Zakharchenya and Korenev 2005).

The magnetic field strength $h_c$ making the magnetic moment of the sample vanish, is referred to as the coercive force. It can be estimated by measuring the zero-field polarization $\rho(H=0)$ after flipping a previously magnetized FM film by an external magnetic field of strength $H^*$. When the switching field $H^*$ is equal to the field $h_c$ the film is demagnetized and the polarization value $\rho(H=0)$ is minimal. The data points on the upper dependence in Fig.3b were measured after switching in the dark. The sharp minimum corresponds to $h_c$=90 Oe. Remarkably, that the coercive force value appeared to be 2.5 times larger than that measured in the same structure with the use of a superconductor quantum interference device (SQUID). The difference between the two techniques is that SQUID registers the total magnetic moment of the structure. It is mainly the magnetic moment of the nickel film whose thickness (*40 nm*) is much larger than that of the interface NiGaAs (a few *nm*). However, the contribution of stray fields of nickel and the interface to the Hanle effect is not reduced to the sum of their magnetic moments. Indeed, Dzhioev et al (1997) have shown that the non-equilibrium electron spin diffuses into n-GaAs over the distance $L_s \approx 10 \mu$ that is 10 times longer than the $1\mu$ size (Bochi et al 1995) of nickel film domain. Therefore the nickel stray fields decay quickly near the surface, and the basic mass of electrons does not "feel" them. On the contrary, the interface stray fields penetrate deep $\geq L_s$ into GaAs and dephase electron spins inside. Therefore, the Hanle effect magnetometer detects the ferromagnetic interface NiGaAs rather than the Ni film itself due to the space selection of their stray fields.

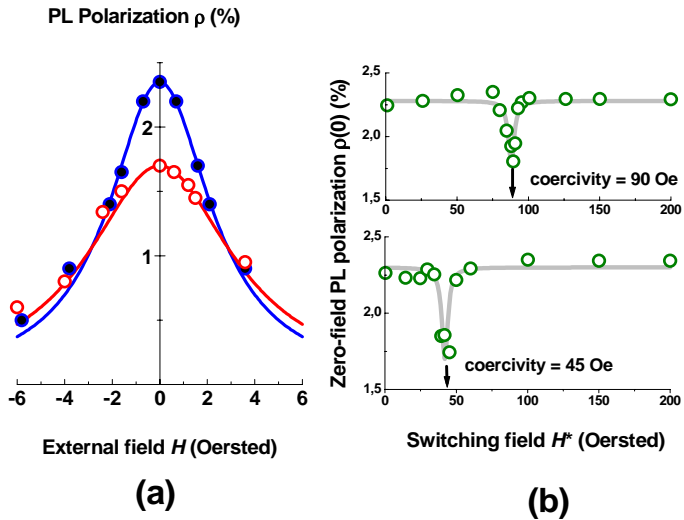

Figure 3. (a) The Hanle effect in a Ni/n-GaAs structure that was previously magnetized (filled circles) in plane and demagnetized (open circles). T=4.2 K.
(b) The degree of zero-field value of circular polarization $\rho(0)$ versus switching field $H^*$. Top panel: switching in the dark, Bottom panel: illumination by He-Ne laser with power density 5 W/cm$^2$ was on during switching process (Adapted from Dzhioev et al 1995).

The bottom curve in Fig.3b was measured under illumination by He-Ne laser. The coercivity of interface decreased twice. This experiment appeared to be the first example of the optical control of ferromagnetism in FM/SC hybrids. It is discussed in detail in section 4.

To detect the small stray field ~*1 Oe* Dzhioev et al (1995) used n-type GaAs where at low temperature (*T=4.2 K*) the halfwidth $H_{1/2} = 2\,Oe$ corresponds to a very long optical orientation lifetime $T_s = 130\,ns$. However the next attempt (Crowell *et al* 1997) using semimagnetic semiconductor ZnCdMnSe (instead of GaAs) with giant g-factor g~100 was unsuccessful because the lifetime $T_s = 15\,ps$ was too short, so that $H_{1/2} \sim 1\,kOe$. Recently Meier et al (2006) have used electrons with longer spin memory $T_s \approx 1.5\,ns$ in InGaAs SC quantum well to successfully observe the stray fields from Fe patterned stripes, where the stray fields were enhanced up to *~200 Oe*.

## 3. Exchange interaction in FM/SC hybrids

*3.1. Exchange interaction of magnetic atoms with semiconductor charge carriers at the FM/SC interface.*

The exchange interaction between magnetic atoms and semiconductor charge carriers is very important in FM/SC hybrids. Tunneling of SC carriers through the interface induces in equilibrium proximity effect – spontaneous ordering of electron spins localized near interface from the semiconductor side. Equilibrium spin polarization decays into the semiconductor over the length of the order of de Broglie wavelength $\lambda_{Br}$. Equilibrium proximity effect is considered in subsection 3.2. A number of new phenomena out of equilibrium are discussed in subsection 3.3.

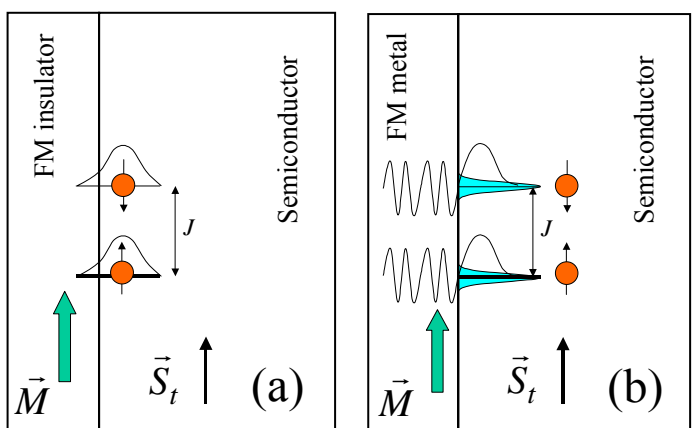

Fig.4. Proximity effect in (a) FM-insulator/SC and (b) FM-metal/SC hybrids. Spin splitting of SC electrons (t-electrons) localized near interface induces their polarization in equilibrium. In case (b) spin sublevels are broadened due to escape into the FM metal.

*3.2. Equilibrium proximity effect.*

The proximity effect in the insulating FM/SC system was considered theoretically by Korenev (1996, 1997). Figure 4a shows a contact between FM insulator and semiconductor with electrons (t-electrons) trapped on deep isolated centers near heterojunction. Tunneling into the ferromagnet leads to their exchange interaction with FM atoms and splits the states of t-electrons with spins along $(\uparrow)$ and opposite $(\downarrow)$ to the magnetization $\vec{M}$. The energy (per unit area) of exchange interaction between t-electrons with surface concentration $n_t$ and mean spin $\vec{S}_t$ and the uniformly magnetized ferromagnet (Appendix) is:

$$E_{ex} = -Jn_t \vec{S}_t \cdot \vec{m} = -\hbar n_t \vec{S}_t \cdot \vec{\omega}, \qquad (3.1)$$

where $J$ is the exchange constant determined by the FM/SC wavefunction overlap[1], $\vec{m}$ is the unit vector along $\vec{M}$. Exchange interaction (3.1) is equivalent to the interaction of each t-electron spin with an effective magnetic field collinear to $\vec{m}$, and vector $\vec{\omega} = J\vec{m}/\hbar$ means the Larmor precession frequency of the t-electron spin in this field. In equilibrium the mean spin polarization of t-electrons at temperature $T$

$$\vec{S}_t = \frac{1}{2} \frac{\vec{\omega}}{\omega} th\left(\frac{\hbar\omega}{2T}\right) \equiv \vec{S}_T \qquad (3.2)$$

is directed along $\vec{\omega}$.

Wavefunctions of t-electrons penetrate deep into the FM in case of the contact with metallic FM film (fig.4b). This induces the broadening of the split $\uparrow$ $(\downarrow)$ states by the $\gamma_+$ ($\gamma_-$) value, respectively. McGuire *et al* (2004) calculated parameters $J$, $\gamma_+$ and $\gamma_-$ for the two-dimensional electron gas near the FM metal for different barriers at the interface. They found that the proximity effect is basically determined by spin-dependent lifetime broadening $\gamma_\pm$, whereas the spin splitting $J$ is sizeable only for a very thin barrier thickness.

Exchange coupling takes place for the contact with p-type semiconductor where the holes are majority charge carriers. Korenev (2003) considered the interaction of the two-dimensional hole gas in quantum well with magnetic atoms of the nearby insulating FM film. The problem is specific in that the size quantization induces the splitting of the $\Delta_{\ell h}$ heavy and light hole states with angular momentum projections $\pm 3/2$ and $\pm 1/2$ onto growth direction $z$ with the heavy holes being the ground state. Merkulov and Kavokin (1995) found that the exchange interaction becomes anisotropic if the splitting $\Delta_{\ell h}$ is larger than the exchange constant $J_h$: $E_{ex} = -J_h n_h S_z^h m_z$. Within this approximation only the out-of-plane magnetization component $m_z \neq 0$ induces the $\pm 3/2$ states splitting. The mean effective spin of heavy holes $\vec{S}^h$ is collinear to the z-axis. Its equilibrium value is given by the equation similar to Eq. (3.2) where the vector $\vec{\omega}$ has only z-component $\omega_z = J_h m_z / \hbar$.

---

[1] Constant J can have any sign in spite of the ferromagnetic ordering inside of FM.

Myers et al (2004) observed the hole spin polarization induced by the proximity effect. The structure consisted of GaAs quantum well between AlGaAs barriers. A ferromagnetic MnAs layer was inserted into one of the barriers. Gate bias controls the overlap of the hole wavefunctions with MnAs. Hole polarization was measured by the degree $\rho$ of the circular polarization of luminescence excited by the linearly polarized light. In the absence of external magnetic field the magnetization $\vec{M}$ lies in the film plane due to demagnetizing field. Deviation of $\vec{M}$ out of plane induces equilibrium hole spin polarization $S_h \leq 6\%$. In wide range of bias the hole spin was opposite to the Mn spin and along $\vec{M}$ in accordance with anticipated antiferromagnetic sign of Mn-hole exchange interaction (Mn g-factor is positive hence magnetization $\vec{M}$ is antiparallel to the Mn spin and parallel $\vec{S}_h$).

*3.3. Optically induced spin polarization of semiconductor electrons near FM/SC interface.*

*3.3.1. Dynamical proximity effect in insulating FM/SC hybrids.*

Optical orientation of t-electrons in the insulating FM/SC (Fig.4a) was considered theoretically by Korenev (1997). The illumination of the hybrid by circularly polarized light with the energy $h\nu$ of quanta greater than the SC energy gap $E_g$ leads to the optical orientation of semiconductor conduction band electrons. It is assumed that the band carriers do not interact with ferromagnet, so that only the t-electrons, localized near interface, undergo the exchange interaction with magnetic atoms. Under these conditions the mean spin $\vec{S}_c$ of the electrons is parallel to the exciting beam and is determined by its helicity. The valence band holes are depolarized due to their fast spin relaxation.

They recombine with t-electrons. The optically oriented conduction band electrons are trapped by the surface states, thus replacing the recombined t-electrons. As a result, the non-equilibrium polarization of t-electrons appears. The steady state mean spin $\vec{S}_t$ obeys the Bloch equation

$$\frac{\vec{S}_c - \vec{S}_t}{\tau_t} + \frac{\vec{S}_T - \vec{S}_t}{\tau_s} + \vec{S}_t \times \vec{\omega} = 0 \qquad (3.3)$$

The first term in Eq.(3.3) describes the generation of spin compensated by recombination with characteristic time $\tau_t$. The second term takes into account the spin relaxation during time $\tau_s$ to equilibrium, whereas the last term gives the precession with Larmor frequency $\vec{\omega}$ around exchange field of magnetic atoms. The solution of Eq.(3.3) is

$$\vec{S}_t = \vec{S}_T \frac{T_s}{\tau_s} + \vec{S}_0 + \frac{\vec{S}_0 \times \vec{\omega} T_s}{1+\omega^2 T_s^2} + \frac{\vec{\omega} \times [\vec{\omega} \times \vec{S}_0] T_s^2}{1+\omega^2 T_s^2} \qquad (3.4)$$

where $\vec{S}_0 = \vec{S}_c \frac{T_s}{\tau_t}$; the optical orientation lifetime $T_s$ is determined by the shortest of the lifetime and the spin relaxation time: $1/T_s = 1/\tau_t + 1/\tau_s$. It follows from Eq. (3.4) that even the non-polarized excitation $(S_0 = S_c = 0)$ decreases the mean spin $S_t$ due to the replacement of polarized equilibrium t-electrons with non-polarized ones. Thus the non-polarized excitation decouples FM and SC. In case of the circularly polarized excitation the mean spin of conduction band electrons $\vec{S}_c$ is not collinear to magnetization $\vec{M}$. Then new components of the t-electron mean spin $\vec{S}_t$ appear: (i) component $\vec{S}_0$ parallel to the conduction band electron spin $\vec{S}_c$, (ii) component $\vec{S}_0 \times \vec{\omega}$ perpendicular to both $\vec{S}_c$ and $\vec{M}$, (iii) component $\vec{\omega} \times [\vec{\omega} \times \vec{S}_0]$ directed from $\vec{S}_0$ to $\vec{M}$. The latter two components result from the precession of non-equilibrium spin of t-electrons in the exchange field of the ferromagnet.

The important role of optically oriented surface states has been demonstrated by Prins et al (1996) in their experiments on the spin-dependent tunneling of spin polarized electrons from GaAs into a ferromagnetic Pt/Co multiplayer sample.

*3.3.2. Dynamical proximity effect in metallic FM/SC hybrids.*

In previous subsection we assumed that the proximity effect is absent for the conduction band electrons. However, if the Schottky barrier is transparent (for example due to the intentional doping of semiconductor) the free electrons reach the interface and undergo the exchange interaction with magnetic atoms. In this case there is noticeable spin-dependent amplitude of transmission of the semiconductor electron into the ferromagnet. Similar to localized states the free electrons acquire the equilibrium spin polarization over the length scale $\sim \lambda_{Br}$ due to the proximity effect with the FM film. However out of equilibrium the conduction band electrons are able to transfer the non-equilibrium spin over the macroscopic distance into semiconductor via drift-diffusion processes. Aronov and Pikus (1976) suggested to inject spin from FM into SC by externally applied bias. The difference in electrochemical potentials of FM $(\mu_f)$ and SC $(\mu_e)$ induces both the particle flux $\vec{q}_e$ and the spin flux (current) due to the spin-dependent tunneling through the heterojunction (Fig.5a). Electrical spin injection was first realized by Alvarado and Reneaud (1992). Recently this research has become very popular (Zutic *et al* 2004).

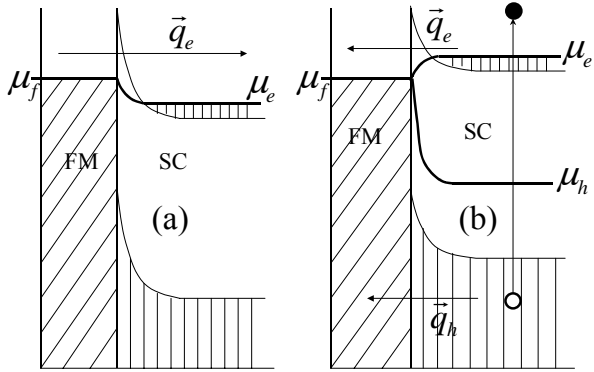

Figure 5. Band diagram of the FM/SC hybrid out of equilibrium due to: (a) electric current; (b) photo-excitation

The electrochemical potential difference can be created optically, too. For example, absorption of non-polarized light creates SC electron-hole pairs whose distribution is determined by the quasi-Fermi levels for electrons $\mu_e$ and holes $\mu_h$ (Fig.5b). The holes are attracted to the FM/SC interface, thus dragging the electrons. Part of the electrons transmits into FM; the other part reflects from the boundary. Bulk non-equilibrium spin polarization of SC electrons develops due to the spin-dependent *extraction* into FM with the polarization sign being opposite to that in the case of injection. The spin density $\vec{S} = n_c \vec{S}_c$ of conduction band electrons with concentration $n_c$ in n-type SC is purely non-equilibrium because $S_T = 0$ in the bulk of non-magnetic SC. The spin dynamics can be described by Bloch equations taking into account diffusion, drift, spin relaxation and Larmor precession in an external magnetic field

$$\frac{\partial \vec{S}}{\partial t} + \frac{\partial \vec{J}^S_\alpha}{\partial x_\alpha} = -\frac{\vec{S}}{\tau_s} + \vec{\Omega} \times \vec{S} \qquad (3.5a)$$

where the vector of spin current in the $x_\alpha$-direction ($x_\alpha = x, y, z$)

$$\vec{J}_\alpha^S = V_\alpha \vec{S} - D \frac{\partial \vec{S}}{\partial x_\alpha} \qquad (3.5b)$$

$V_\alpha$ is the $\alpha-$component of drift velocity, $D$ – diffusion coefficient. The efficiency of spin injection/extraction is determined by the processes at the FM/SC interface and is taken into account in the boundary condition for the normal component $\vec{n}$ of spin current as proposed by Aronov and Pikus (1976)

$$\vec{J}_n^S = a \cdot \vec{m} \qquad (3.6)$$

where the parameter $a$ is determined by the spin-dependent transmission/reflection at the interface. Parameter $a$ depends on the applied bias in the case of electrical injection (Fig.5a) and by the generation/recombination processes for the optical injection (Fig.5b). In any case the SC electrons acquire the non-equilibrium spin density $\vec{S} \propto \vec{m}$, which can be transferred into semiconductor.

Optically induced spin polarization of SC electrons was observed by Kawakami et al (2001) in MnAs/n-GaAs and GaMnAs/n-GaAs heterostructures. Under illumination by linearly polarized light electrons acquire a non-equilibrium spin density $\vec{S}$, whose direction is collinear to the magnetization $\vec{M}$ of nearby FM. Vector $\vec{M}$ lies in the plane due to the demagnetization field, which excludes the possible role of the magnetic circular dichroism (MCD) effect. The spin density $\vec{S}$ inside a semiconductor undergoes Larmor precession around an external magnetic field $\vec{H}$. Another evidence of the accumulated non-equilibrium spin results from the dynamical nuclear polarization. It is known (Meier and Zakharchenya Eds. 1984) that out of equilibrium electrons transfer their angular momentum into the nuclear spin-system due to the hyperfine interaction. In turn, polarized nuclei create an effective hyperfine magnetic field changing the precession frequency of electrons in an external field. Kawakami et al (2001) observed this effect referred to as "ferromagnetic imprinting of nuclear spins". Epstein et al (2003) studied an optically induced dynamical proximity effect combined with an externally applied bias. They found that the value of the optically induced spin density $\vec{S} \propto \vec{m}$ varies strongly with bias: it is maximal at a forward bias. This result can be understood from Fig.5b. Forward bias (+ at ferromagnet) rectifies the band bending, thus increasing the electron spin current through the FM/SC interface. It results in the increased spin density.

*3.3.3. Dynamic proximity effect for the non-collinear spins of SC and FM at the interface*

Consider the case when the spin density $\vec{S}$ near the FM/SC boundary is non-collinear to magnetization $\vec{M}$. It appears, for example, under optical orientation conditions, i.e. excitation by circular polarized light (Fig.1). Optically oriented electrons reach the FM/SC boundary, partly going through (reflecting from) it with the change of their initial spin orientation. This problem reminds the well-known problem of scattering of the spin-polarized electron beam by magnetic target (Kessler 1985). Besides the components $\vec{m}$ and $\vec{m}(\vec{m}\cdot\vec{S})$ along magnetization, the reflected electrons acquire transverse components $\vec{m}\times\vec{S}$ and $\vec{m}\times(\vec{m}\times\vec{S})$, which describe the rotation of spin $\vec{S}$ and its relaxation toward $\vec{M}$, respectively. It is reasonable to apply the old methods exploiting the matrix of density to the FM/SC problem. Ciuti et al (2002) calculated the spin-dependent reflection coefficients for the FM/SC heterojunction within the effective mass approach. A little bit earlier Brataas et al (2001) considered spin the transport through the paramagnet metal-FM junction for non-collinear polarizations. They deduced the boundary condition for the spin current

$$\vec{J}_n^S = a\cdot\vec{m} + b(\vec{m}\cdot\vec{S})\vec{m} + c[\vec{m}\times\vec{S}] + r[\vec{m}\times[\vec{m}\times\vec{S}]] \tag{3.7}$$

where the parameters $a$ and $b$ are determined by the spin-dependent conductance $g_\uparrow$ ($g_\downarrow$) for the electron spins along (oppose) to the FM magnetization, parameter $c$ ($r$) is given by imaginary (real) part of the so-called mixing conductance ($g_{\uparrow\downarrow}$). The third term describes the spin precession around an effective exchange magnetic field, whereas the forth term gives the relaxation of $\vec{S}$ toward $\vec{M}$. They are similar to the last two terms in Eq.(3.4) for the insulating FM/SC hybrid (subsubsection 3.3.1).

Strictly speaking not only the spin current but also the charge current through the FM/SC boundary is spin-dependent. Johnson and Silsbee (1985) argued that the electron current at the FM-paramagnetic metal interface

$$j_e = (g_\uparrow + g_\downarrow)\cdot[\Delta\xi + \alpha(\vec{m}\cdot\vec{S})] \tag{3.8}$$

is determined not only by the electrochemical potential difference $\Delta\xi$ but also by the projection of $\vec{S}$ onto $\vec{M}$.

An experimental study of the effects given by the first term in Eq. (3.7) was discussed in subsubsection 3.3.2. To the best of my knowledge, the experimental observation of the other three terms in Eq.(3.7) has not been reported yet. Therefore, the experimental proof of the Eq. (3.7) (and Eq.(3.4) in insulating hybrids) for the non-collinear spins remains a challenge for experimentalists.

The spin-dependent charge current of optically oriented electrons Eq.(3.8) was studied by Prins et al (1996), Isakovic et al (2001) and Hirohata et al (2001). The authors modulated spin density $\vec{S}$ of SC electrons via modulation of the helicity of photons. A magnetic field perpendicular to the structure plane brought the magnetization out of the film plane and induced the non-zero projection $(\vec{S}\cdot\vec{m}) \neq 0$. The signal of photocurrent through the interface was detected. At the same time Prins et al (1996), Isakovic et al (2001) stated the strong influence of the MCD effect, i.e. the difference in absorption coefficients $\Delta\alpha \propto (\vec{S}\cdot\vec{m})$ of the left and right circular polarizations. Hence care should be taken for the quantitative interpretation of the results.

### 3.3.4. Dynamics of dynamic proximity effect

Epstein et al (2002) observed the two-step dynamics of the non-equilibrium spin polarization of SC electrons due to the dynamic proximity effect. The electron spin polarization appears during the first 50 ps after excitation by linearly polarized laser pulse – the first step. The second step is the long-term spin relaxation of the order of nanoseconds. Bauer et al (2004) carried out calculation of the two-step dynamics for the case of uniform spatial distribution of electron spin inside of SC, when drift-diffusion length is larger than the SC thickness. They found that the first step results from the fast withdrawal of photo-holes toward the interface under the action of a built-up electric field (Fig.5b). In turn, this induces the spin-dependent transfer of electrons into the ferromagnet (spin extraction) accompanied by the accumulation of non-equilibrium spin in semiconductor. The non-equilibrium electron spin survives tens of nanoseconds at low temperature in n-type GaAs based semiconductors (Dzhioev et al 1997). This explains the long-term stage of the spin dynamics. Gridnev (2007) calculated the nanosecond dynamics with the use of Eqs.(3.5, 3.7) for the case of thick semiconductor when the spin density $\vec{S}$ is spatially non-uniform. He found the non-uniform distribution of spin density due to the Larmor precession and spin relaxation on its way into the interior of semiconductor.

## 4. Optical control of ferromagnetism in FM/SC hybrid.

It is well known that light, besides trivial heating, exerts non-thermal influence upon the magnetic properties of ferromagnetic semiconductors (Nagaev 1988), such as Curie temperature, magnetic hysteresis loop, magnetic anisotropy etc. Recently similar phenomena have been reported in new GaAs based ferromagnetic semiconductors (Wang et al 2005). This section considers the physics underlying the optical control of ferromagnetism specifically for the FM/SC hybrids: magnetism of the hybrid (magnetic hysteresis loop and the orientation of magnetization vector in space) is tuned optically with the help of semiconductor. The electrical manipulation of ferromagnetism was considered by Zakharchenya and Korenev (2005).

*4.1. Increase of the exchange stiffness coefficient of the FM/SC with electron accumulation layer.*

In case of an electron accumulation layer the exchange interaction of magnetic atoms with SC electrons is isotropic. Korenev (1996) considered the exchange interaction between magnetic atoms and SC electrons localized at deep centers near the FM/SC interface (subsection 3.2, Fig.4a). Exchange favors collinear orientation of t-electron spin and the magnetic atom spins located within the localization area of the t-electron. The local enhancement of the ferromagnetic ordering increases the exchange rigidity coefficient of the hybrid in equilibrium.

Quantitative analysis is based on the Eq.(3.1) that should be modified in the non-uniform case (Korenev 1996, 1997). Assume that magnetization $\vec{m}(x)$ changes orientation in one direction $x$ over the distance $\delta$ (domain wall thickness) much larger then the localization radius $a_t$. Within the macroscopic approach (see Appendix) the exchange energy per unit surface area

$$E_{ex}(x) = -\hbar n_t(x)\vec{S}_t(x)\cdot\vec{\omega}(x) \tag{4.1}$$

where the frequency

$$\vec{\omega}(x) = \frac{J}{\hbar}\left[\vec{m}(x) + \frac{a_t^2}{2}\vec{m}_x''(x)\right] \quad (4.2)$$

contains in comparison with Eq.(3.1) an additional term with the second space derivative. It results from the smooth variation of $\vec{m}(x)$ within the localization area $\sim \pi a_t^2$ and represents the first nontrivial term of expansion of the exchange in series in powers $(a_t/\delta)^2 \ll 1$. As a result the mean spin of t-electrons also depends on $x$: Eq.(3.2) gives its equilibrium value with vector $\vec{\omega}(x)$ being taken from Eq.(4.2). At low temperature the mean spin $\vec{S}_t(x) = \vec{\omega}(x)/2\omega(x)$. Substituting the $\vec{S}_t(x)$ and Eq. (4.2) into Eq. (4.1) one finds

$$E_{ex}(x) = \varepsilon_0 + \frac{A_1}{2}\left(\frac{\partial \vec{m}}{\partial x}\right)^2 \quad (4.3)$$

where $\varepsilon_0 = -|J|n_t/2$ does not depend on the distribution of magnetization in space; the exchange stiffness coefficient $A_1 = |J|n_t a_t^2/2$; the identity $\vec{m}\cdot\vec{m}'' = -(\vec{m}')^2$ was used. Eq.(4.3) shows that the exchange interaction increases the exchange stiffness of the hybrid by $A_1$ value in agreement with qualitative discussion.

The increase of the exchange stiffness coefficient gives an additional contribution to the coercivity of the hybrid. However let us recall first the origin of coercivity.

*4.2. Origin of coercivity (Chikazumi 1984).*

The width of the magnetic hysteresis loop is often determined by the displacement of the domain walls separating adjacent domains. Space fluctuations of the anisotropy constant $\Delta K(x)$ or the exchange stiffness coefficient $\Delta A(x)$ about their average values $K_0$ and $A_0$ induce the fluctuations of the domain wall energy $\gamma(x) \sim \sqrt{[A_0 + \Delta A(x)]\cdot[K_0 + \Delta K(x)]}$. The potential relief $\gamma(x)$ fixes the domain walls giving the source of coercivity. Coercivity $h_c$ is determined by both the amplitude of fluctuations $\Delta\gamma$ and their characteristic length $\Lambda$. An external magnetic field $H$ exerts pressure $P = 2MH$ on $180^0$-domain wall. It is compensated by the pressure $\frac{\partial \gamma(x)}{\partial x} \sim \frac{\Delta\gamma}{\Lambda}$ due to the fluctuations. The coercivity can be estimated by equating the pressure at $H = h_c$ to the maximum restoring force $2Mh_c = \left.\frac{\partial \gamma(x)}{\partial x}\right|_{max} \sim \frac{\Delta \gamma}{\delta}$. Here we have taken into account that the maximum is reached when $\Lambda$ is close to the domain wall thickness $\delta = \sqrt{A_0/K_0}$: pinning is negligible at $\Lambda \gg \delta$, whereas the effect of fluctuations is averaged out over the wall thickness at $\Lambda \ll \delta$. We conclude that one should decrease the amplitude of potential relief to decrease coercivity. For small fluctuations $(\Delta \gamma \ll \gamma)$ we estimate $h_c \sim \Delta K/M$ for the fluctuations of the anisotropy constant and $h_c \sim \Delta A/(M \cdot \delta^2)$ for the fluctuations of the exchange stiffness coefficient.

*4.3. Optical control of coercivity of the FM/SC with electron accumulation layer.*

In case of a two-dimensional electron accumulation layer the exchange of magnetic atoms with SC electrons is isotropic and affects the exchange stiffness coefficient $A_1$ (subsection 4.1). The fluctuations $\Delta n_t$ of the surface density $n_t(x)$ of t-electrons create fluctuations $\Delta A_1 = |J|\Delta n_t a_t^2/2$. If $\Delta n_t \sim n_t$ then the coercivity is enhanced in the dark by the $\delta h_c(0) \sim A_1/M_{surf}\delta^2$ value ($M_{surf} = M \cdot d$ is the magnetic moment per unit surface area, $d$ is the width of the FM film). Under illumination of the hybrid by light with power density $W$ and photon energy $h\nu > E_g$ the photo-excited holes move to the interface (Fig.5b) and recombine with t-electrons, thus decreasing both $n_t$ and coercivity (Korenev, 1996)

$$h_c(W) = h_1 + \frac{\delta h_c(0)}{1 + W/W_0} \qquad (4.4)$$

where $W_0$ is the characteristic power density depending on the absorption coefficient and the capture efficiencies of electrons and holes by t-centers, parameter $h_1$ takes into account other contributions to coercivity.

The optical control of the coercivity of FM/SC hybrid has been demonstrated experimentally by Dzhioev et al (1994, 1995) in a Ni/n-GaAs structure. We have already considered this system in subsection 2.2 and found that the optically oriented electrons represent a sensitive detector of weak stray fields of the Ni-GaAs interface rather than Ni film itself. The important feature of the Ni/n-GaAs hybrid consists in the optical tunability of the interface ferromagnetism. Figure 6 shows that the illumination by He-Ne laser ($h\nu = 1.96\,\text{eV}$) decreases the interface coercivity by half (compare also the lower dependence on Fig.3b with the upper one). However illumination does not affect the coercive force of nickel film. The effect we called *photocoercivity* is not sensitive to the light polarization and takes place only in the illuminated region. Photocoercivity takes place at a low power density (a few mW/cm$^2$) and is not related to the heating of sample: the heating of the hybrid by passing the dc electric current across the FM/SC heterojunction, with dissipated power being 10 times larger than the power of light, remained the coercive force unchanged. Spectral measurements have shown that the photocoercivity is due to the effect of semiconductor on ferromagnetic interface: it diminishes if the photon energy is below the energy gap of GaAs. The solid line in Fig.6 is calculated with the use of Eq.(4.4) for $h_1 = \delta h_c(0) = 45\,Oe$, $W_0 = 8\,mW/cm^2$ showing good agreement.

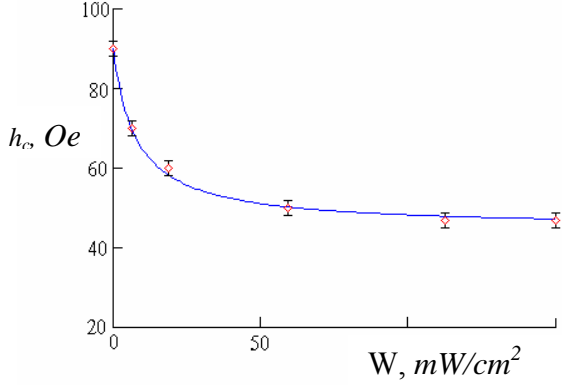

Fig.6. Dependence of the interface coercivity on light intensity. Adapted from Dzhioev et al 1995.

The question about the origin of the interface ferromagnetism remains for future studies. Lahav et al (1986) have shown that already at $T \geq 100\,^0C$, Ni diffuses into GaAs to form an intermediate layer. This layer may be ferromagnetic and form what we understand by "interface". However, I have no information on the research in magnetism of NiGaAs compounds.

*4.4. Surface anisotropy in the FM/SC with hole accumulation layer.*

Another situation arises from the contact of the FM/SC with a hole accumulation layer. For example, it takes place in inversion layers, with strong band bending (Fig.7), or in case of the p-type quantum well near FM. Size quantization leads to anisotropic exchange interaction (subsection 3.2) with the energy per unit area

$$E_{ex} = -J_h n_h S_z^h m_z = -\hbar n_h S_z^h \omega_z \qquad (4.5)$$

and frequency $\omega_z = J_h m_z/\hbar$. At low temperature the holes are completely spin-polarized due to proximity effect (subsection 3.2) with mean spin $S_z^h = \omega_z/2|\omega_z|$. Then the exchange energy $E_{ex} = -\frac{|J_h m_z|}{2} n_h$. One can see that the exchange interaction can be considered as a peculiar magnetic surface anisotropy with easy axis being directed along z-axis (perpendicular anisotropy). This is in striking contrast with the case of an electron accumulation layer (subsection 4.1). If the exchange constant and hole concentration are strong enough, then the orientational transition is possible (Korenev 2003): the magnetic moment leaves the plane and becomes oriented along the normal.

*4.5. Photocoercivity in the FM/SC with the hole accumulation layer.*

Similar to subsections 4.2, 4.3 the fluctuations of exchange coupling between magnetic atoms and SC holes induce fluctuations of the anisotropy constant $\delta K \sim J_h n_h$ increasing the coercivity in the dark by $\delta h_c(0) \sim J_h n_h / M_{surf}$. The fluctuations $\delta K$ can be much less than the averaged anisotropy constant $K_0$, and no visible change of the average magnetic anisotropy takes place. However they are crucial for the coercivity. The illumination of the hybrid may decrease the exchange constant $J_h$ due to the band flattening, thus decreasing the overlap of the hole wavefunctions with magnetic atoms (Fig.7).

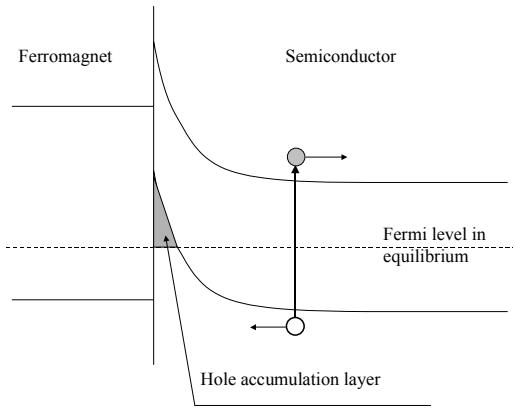

Fig.7 Band diagram of the FM/SC hybrid with hole inversion layer

Oiwa et al (2001) demonstrated the optical control of the coercive force of p-InMnAs FM grown on non-magnetic GaSb semiconductor. The authors observed that the photo-excitation of the hybrid decreased the coercive force. This effect was the strongest under the illumination by light with the photon energy larger than the energy gap of nonmagnetic semiconductor GaSb, thus pointing to the important role of SC. Strong bend bending in GaSb (similar to the Fig.7) takes place near the FM/SC interface in the dark. This leads to accumulation of a substantial number of holes in the vicinity of interface. Authors explained the photocoercivity as due to the change of ferromagnetism by the holes excited in the GaSb and transferred into the InMnAs layer. Another explanation follows from the above discussion: one could take into account the weakening of the FM/SC exchange coupling near interface.

*4.6. Photo-induced change of the Curie temperature.*

Light can affect the Curie temperature $T_C$, too. For example, the photoexcitated carriers may change the exchange constants of ferromagnets. The light-induced increase in Curie temperature $\Delta T_C$ is well known in ferromagnetic semiconductors (Nagaev 1988). A simple estimation shows that this effect is pretty small: $\Delta T_C \approx (n/N_{fm}) T_C \approx 0.1 K$ if the concentration of magnetic atoms $N_{fm} \sim 10^{22} \, cm^{-3}$, $T_C = 1000 \, K$ and the concentration of photocarriers $n = 10^{18} \, cm^{-3}$.

It is reasonable to expect a similar small change in the FM/SC hybrids. However, Koshihara et al (1997) claimed the photo-induced ferromagnetic order in p-InMnAs/GaSb hybrid persistent up to 35 K. The authors measured the magnetization component normal to the sample plane vs magnetic field with the use of SQUID and anomalous Hall effect. They found non-hysteretic behavior of magnetization in the dark. In contrast, excitation with photon energy larger than the energy gap of GaSb induced clear hysteresis. The authors interpreted this effect as a paramagnet-ferromagnet transition of magnetic semiconductor InMnAs due to the transfer of SC photo-holes into InMnAs FM. In other words, the change $\Delta T_C \approx 35 K$ takes place under the change $\Delta n_h \approx 10^{18} \, cm^{-3}$ in hole concentration inside of InMnAs (from 3.76*$10^{19}$ cm$^{-3}$ in the dark up to 3.90*$10^{19}$ cm$^{-3}$ on the light). It is not clear whether the existing theories (for example, Dietl et al 2000, Kaminski and Das Sarma 2003) are able to explain this unusual effect.

*4.7. Effect of circularly polarized light on the magnetization.*

The circularly polarized light transfers the angular momentum $\pm \hbar$ per photon. Hence its absorption magnetizes the sample. For example, it excites spin oriented charge carriers (Meier and Zakharchenya 1984). Ordinarily their concentration is much smaller than that of magnetic atoms, so that the direct magnetization by light is inefficient. Much stronger effect consists in the appearance of an effective magnetic field proportional to the degree $P_c$ of circular polarization of light. If the field value is larger than coercivity, then the sample is magnetized even in the absence of external magnetic

Optical orientation in ferromagnet/semiconductor hybridsfield. Van der Ziel et al (1965) proposed inverse Faraday effect: the circularly polarized light acts on atoms in *non-absorbing* media as a magnetic field lifting Kramer's degeneracy. This field is relatively weak: $H_{eff}$ ~0.01 Oe at power density $10^7$ W/cm$^2$ for Er$^{+2}$:CaF$_2$. Therefore optical pulses heating the magnetic system close to TC should perform the magnetization. Stanciu et al (2007) reported the optical magnetization of the ferrimagnetic GdFeCo.

Effective magnetic fields can be created under adsorption of the circular polarized light. In this case optically oriented carriers undergo a strong exchange interaction with magnetic atoms. Merkulov and Samsonidze (1980) considered theoretically the domain wall motion under the action of the circularly polarized light exciting the ferromagnetic semiconductor with perpendicular anisotropy. The magnetic circular dichroism (MCD) effect leads to the optical orientation of carriers: a larger electron concentration is excited in domains with one orientation of $\vec{M}$ with respect to the domains with the opposite $\vec{M}$. Then the size of one type of domains increases at the expense of the others. Merkulov and Samsonidze (1980) solved the dynamic Landau-Lifshitz equation and found that the domain wall moves as if it was an effective magnetic field whose value and direction is determined by the helicity of light. Nagaev (1988) reviewed early experiments in ferromagnetic semiconductors on this topic.

*4.7.1. Effective magnetic fields in FM/SC with electron accumulation layer.*

Korenev (1997) considered theoretically the action of circularly polarized light on the domain wall in the FM/SC hybrids. He argued that the effective field $\vec{H}_{eff}$ appears as a result of a pressure exerted by the optically oriented semiconductor electrons on the domain wall. Both the usual optical orientation of carriers in non-magnetic semiconductor and the MCD effect result in the appearance of the effective field whose value and direction are determined by the angular momentum $\hbar \cdot \vec{P}_c$ of the photon. Imagine a thin FM film with perpendicular anisotropy grown on the semiconductor surface with distribution $\vec{M}(x)$ in a 180$^0$-wall varying in one direction $x$ (Fig.8). Below we shall assume the usual distribution of $\vec{m}(\vec{r})$ within the domain wall $m_z(x) = \cos\theta(x)$, $m_y(x) = \sin\theta(x)$, $m_x(x) = 0$, $\cos\theta(x) = -th(x/\delta)$ (Chikazumi 1984). Let the exchange interaction between magnetic atoms and SC t-electrons be small enough not to affect both the domain wall distribution and thickness $\delta$. The domain wall centered at the position $x = x_0$ undergoes the pressure due to the exchange interaction between magnetic atoms and SC electrons

$$P = 2M_{surf} H_{eff} = -\frac{\partial}{\partial x_0}\left[\int_{-\infty}^{+\infty} E_{ex}(x - x_0)dx\right] = \hbar \int_{-\infty}^{+\infty} n_t \vec{S}_t \frac{\partial \vec{\omega}(x - x_0)}{\partial x_0} dx \qquad (4.6)$$

where the exchange energy per unit surface area $E_{ex}(x)$ and $\vec{\omega}(x)$ are given by Eq. (4.1) and Eq. (4.2), respectively. Space derivation in Eq.(4.6) should be performed under fixed $n_t \vec{S}_t$ in spite of the fact that it may (and really do) depends on $\vec{m}(x)$ distribution. This is because the calculation of force in open systems should be performed under fixed external conditions (Landau and Lifshits 1979). In our case the spin system of t-electrons is an external system with respect to the ferromagnetic one.

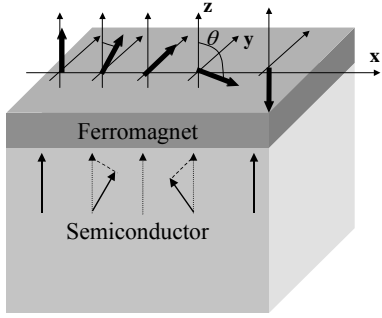

Fig.8 Bold arrows show the magnetization profile within the 180-wall. Thin solid arrows show the projection of the initial spin (up) onto vector $\vec{\omega}$

In equilibrium the force will be absent. Indeed, the substitution of equilibrium value $\vec{S}_T$ Eq.(3.2) into Eq.(4.6) gives zero. Hence the nonequilbrium spin density of t-electrons $n_t \vec{S}_t$ should be calculated. In the MCD case the mean electron spin is in equilibrium and completely spin-polarized at low temperature, whereas the concentration of t-electrons is different in different domains

$$n_t(x) = n_t(1 + \gamma \vec{P}_c \cdot \vec{m}(x)), \quad \vec{S}_t(x) = \vec{\omega}(x)/2\omega(x) \qquad (4.7)$$

Parameter γ characterizes both the value and sign of dichroism. Substituting Eq.(4.7) into Eq.(4.6) we reproduce Merkulov and Samsonidze (1980) result deduced from the Landau-Lifshitz dynamic equation

$$H_{eff} = \gamma P_c \frac{A_1}{3M_{surf} \delta^2} \qquad (4.8a)$$

Alternatively one may consider the case of optically oriented electrons when the non-equilibrium spin $\vec{S}_t$ is governed by the equation (3.4) but the concentration $n_t = const(x)$ is the same in different domains (no MCD effect). Taking into account that only the projection $(\vec{S}_t \cdot \vec{\omega})\vec{\omega}/\omega^2$ of the spin $\vec{S}_t$ of t-electrons is conserved in the case of strong exchange $\omega T_s \gg 1$ we obtain from Eqs.(3.4, 4.6)

$$H_{eff} = 2S_c \frac{T_s}{\tau_t} \frac{A_1}{3M_{surf}\delta^2} \qquad (4.8b)$$

Note that the expressions (4.8) for the field $\vec{H}_{eff}$ contain a small parameter $a_t^2/\delta^2 \sim 10^{-2} - 10^{-3}$ (for typical values of localization radius $a_t \sim 1nm$, and domain wall thickness $\delta \sim 30nm$), so that for the point defect $H_{eff} = 0$. This is the result of the flexibility of the semiconductor electron spin system adjusting the direction of the mean spin to $\vec{M}$.

*4.7.1. Effective magnetic fields in FM/SC with hole accumulation layer.*

The effective field value increases drastically if the direction of t-electron spin is fixed, thus the spin precession in the ferromagnet exchange field is absent. Suppose that a p-type accumulation layer with hole concentration $n_h$ is formed in semiconductor (or quantum well filled with holes) near the FM/SC interface. Due to the reduced symmetry the hole spin states are split into heavy and light hole doublets as discussed in subsections 3.2, 4.4. Then the hole spin $S_h = m_z/2$ is fixed along z-direction. Taking into account the expression for the Larmor frequency $\omega_z = J_h m_z(x)/\hbar$ and using Eq.(4.6) we get for the MCD case

$$H_{eff}^h = \gamma P_c \frac{J_h n_h}{6M_{surf}} \qquad (4.10a)$$

and for the case of the heavy hole optical orientation

$$H_{eff}^h = S_h \frac{T_s}{\tau} \frac{J_h n_h}{M_{surf}} \qquad (4.10b)$$

One can see that the $H_{eff}$ field is indeed greatly ($\delta^2/a_t^2$ times) enhanced due to the rigidity of the hole spin system.

The effective field will shift the magnetic hysteresis loop *M(H)* by the $H_{eff}$ value over the *H* axis. Let us estimate the $H_{eff}$ field value for the FM/SC coupling constant $J_h = 0.1\ eV$ and the hole surface concentration $n_h = 10^{12}\ cm^{-2}$. We take $M_{surf} \approx 2\mu_B N_{fm} d \approx 2 \cdot 10^{15}$ Bohr magnetons per unit area for the FM film thickness $d = 1\,nm$ and the concentration of FM atoms $N_{fm} \approx 10^{22}\,cm^{-3}$. For the MCD case we find from the Eq.(10a) that $H_{eff} = 1400\,Oe$ under favorable condition $\gamma = 1$ and $P_c = 1$ for the 100 % circularly polarized light. For the case of the optical orientation of holes we get for $2 S_c T_s / \tau_t = 1$ a value 4200 Oe. We conclude that the exchange coupling with holes looks very promising for the optical control of ferromagnetism of the hybrid.

Oiwa et al (2002) observed the magnetization of GaMnAs FM film by the circularly polarized light in the GaMnAs/GaAs hybrid. The authors explain it by the photo-creation inside GaMnAs of spin-oriented holes, which dynamically polarize the Mn spins. However, a very sharp spectral dependence correlates with the excitation of paramagnetic GaAs rather than ferromagnetic GaMnAs whose spectrum is very smooth due to a strong disorder. This fact provides a strong evidence of the crucial role of GaAs excitation in the optical magnetization of GaMnAs. Therefore, one could also consider the possible role of optically oriented holes in GaAs and their exchange with magnetic atoms as discussed in this section.

## 5. Summary

Spin-spin interactions in the FM/SC hybrid lead to a strongly coupled spin system of ferromagnet and semiconductor. On the one hand they induce the proximity effect – spin polarization of semiconductor electrons. Hence semiconductor electrons monitor the magnetic state of the ferromagnet. On the other hand the magnetic properties of the unified system differ drastically from those of the FM film alone. The magnetism of the entire system can be controlled optically. As a result the hybrid constitutes an elementary magnetic storage with the semiconductor being not only a substrate but an active participant in information processing. An additional degree of freedom consisting in the choice of desirable FM/SC pair among paramagnet semiconductors and a large number of ferromagnetic materials provides many possibilities. The ultimate goal is the discovery of FM/SC hybrids with the optical control of magnetism at room temperature. The most promising for this purpose seems to be the FM/SC system operating on the optically tunable proximity effect.

Author greatly appreciates I.A. Merkulov for valuable remarks. The paper is supported in part by RFBR, Russian Science Support Foundation and programs of Russian Academy of Sciences.

**Appendix**

Equations (3.1, 4.1) can be derived from the Hamiltonian $\hat{H} = \dfrac{J}{N_{fm}I}\sum_{i,j}\hat{\vec{s}}_i \cdot \hat{\vec{I}}_j \delta(\vec{r}_i - \vec{R}_j)$ describing the isotropic exchange interaction between t-electrons and magnetic atoms with spin $I$ and concentration $N_{fm}$. Here $\hat{\vec{s}}_i$ and $\hat{\vec{I}}_j$ are the operators of spins of the i-th t-electron and j-th magnetic atom located at $\vec{r}_i$ and $\vec{R}_j$, respectively. Deep centers are assumed to be isolated ($n_t a_t^2 \ll 1$).

Assume that the localization radius of t-electron $a_t$ is much larger than the distance $a_0$ between magnetic atoms. In this case ferromagnet can be considered as a continuous medium with macroscopic magnetization $\vec{M}(\vec{r})$. Replacing the spin density operator $\hat{\vec{I}}(\vec{r}) = \sum_j \hat{\vec{I}}_j \delta(\vec{r} - \vec{R}_j)$ with its classical value $\vec{I}(\vec{r}) = -IN_{fm}\vec{m}(\vec{r})$ (the sign is negative because the spin of the magnetic atom is usually antiparallel to its magnetic moment) we arrive at the mean-field Hamiltonian $\hat{H} = -J\sum_i \hat{\vec{s}}_i \cdot \vec{m}(\vec{r}_i)$. It describes the interaction of each electron spin with an effective non-uniform magnetic field $\propto \vec{m}(\vec{r})$. If the exchange interaction acts as perturbation we obtain the spin Hamiltonian

$$\hat{H}_S = -J\sum_i \hat{\vec{s}}_i \int \Phi_i^2(\vec{r})\vec{m}(\vec{r})d\vec{r} \tag{A1}$$

by averaging the mean-field Hamiltonian with the ground state orbital wavefunction $\Phi_i(\vec{r})$ of t-electron with coordinate $\vec{r}$ located near the *i*-th center. It is reasonable to consider the magnetization to be a slowly varying function, because the characteristic distance $\delta \approx 30\,nm$ is much larger than $a_t \approx 1\,nm$. Then the integral in Eq.(A1) can be expanded in series in powers $(a_t/\delta)$. If the magnetization $\vec{m}(x)$ varies only along *x*-direction then we have up to the second order terms

$$\hat{H}_S = -J\sum_i \hat{\vec{s}}_i \cdot \left(\vec{m}(x_i) + \dfrac{a_t^2}{2}\vec{m}''(x_i)\right) \tag{A2}$$

Here magnetization and its second derivative are taken at the position $x_i$ of the *i*-th center. Localization radius is determined by the relation $a_t^2 \equiv \int (x-x_i)^2 \Phi_i^2(\vec{r})d\vec{r}$. Equation (A2) has a form

Optical orientation in ferromagnet/semiconductor hybrids

$\hat{H}_S = \hbar \sum_i \hat{\vec{s}}_i \cdot \vec{\omega}_i$, where the vector is given by Eq.(4.2) and summing goes over all paramagnetic centers. Introducing the spin density operator of t-electrons $\hat{\vec{S}}(\vec{r}) \equiv \sum_i \hat{\vec{s}}_i \delta(\vec{r} - \vec{r}_i)$ we can present it in equivalent form

$$\hat{H}_S = \hbar \int \vec{\omega}(\vec{r}) \cdot \hat{\vec{S}}(\vec{r}) d^2 \vec{r} \qquad (A3)$$

We can further treat the problem macroscopically if the number of centers within the characteristic area $\geq \delta^2$ is very large: $n_t \delta^2 \gg 1$. Replacing in this case the spin density operator $\hat{\vec{S}}(\vec{r})$ by its classical value $n_t(\vec{r}) \vec{S}_t(\vec{r})$ we arrive at the quasiclassical expression for the total interaction energy

$$E_{Total} = \int E_{exc}(\vec{r}) d^2 \vec{r} = \hbar \int n_t(\vec{r}) \vec{S}_t(\vec{r}) \cdot \vec{\omega}(\vec{r}) d^2 \vec{r} \qquad (A4)$$

This justifies the Eq.(4.1) for the exchange energy per unit area $E_{exc}(\vec{r})$, which reduces to the Eq.(3.1) for the uniform $\vec{M}$.

The limits of applicability of the macroscopic description bring about the inequality $1/a_0^2 \gg 1/a_t^2 \gg n_t \gg 1/\delta^2$. For $a_0 = 3 \cdot 10^{-8}$ cm, $a_t = 10^{-7}$ cm and $\delta = 3 \cdot 10^{-6}$ cm it is satisfied in the wide concentration range $n_t \in [10^{11} ... 10^{14}]$ cm$^{-2}$.